\documentclass[epj]{webofc}
\usepackage[varg]{txfonts}   
\wocname{International Conference on New Frontiers in Physics} 
\woctitle{International Conference on New Frontiers in Physics 2013}
\begin{document}
\title{Kaon femtoscopy of Pb-Pb and pp collisions at the LHC with the ALICE experiment. }
%
%


\author{L. V. Malinina for the ALICE Collaboration \inst{1,3}\fnsep\thanks{\email{lmalinin@mail.cern.ch
    }} 
}

\institute{Joint Institute for Nuclear Research, Dubna, Russia  
\and
           Skobeltsyn Institute of Nuclear Physics, Lomonosov Moscow State University, Moscow, Russia  
\and
           Moscow, Russia
          }

\abstract{%
 We report on the results of femtoscopic analyses of Pb-Pb collisions at 
$\sqrt{s_{NN}}$=2.76~TeV and pp collisions at $\sqrt{s}$= 7~TeV 
with identical charged and neutral kaons. 
The femtoscopic correlations allow one to measure the space-time
characteristics of particle production using particle correlations due to 
the effects of quantum statistics for identical particles and final state 
interaction for both identical and non-identical ones. Small contributions 
from resonance decays make kaon femtoscopy an ideal tool for the correlation 
studies. In conjunction with pion and proton femtoscopy,
they can also reveal properties of collective dynamics in 
heavy-ion collisions. ALICE results are compared with the
existing world data on kaon femtoscopy in different type of collisions. 
The theoretical expectations for pp and Pb-Pb collisions are considered
}
\maketitle
\section{Introduction}
\label{intro}

Extremely high energy densities achieved in heavy-ion collisions at the Large Hadron Collider (LHC) may entail the formation of the Quark-Gluon Plasma (QGP), a state characterized by partonic degrees of freedom  \cite{QGP}. Studying the QGP is the main goal of the ALICE experiment
(A Large Ion Collider Experiment) \cite{ALICE1}. The highly compressed strongly-interacting system is expected to undergo longitudinal and transverse expansions.
Experimentally, the expansion and the spatial
extent at decoupling are observable via Bose-Einstein correlations. 
Bose-Einstein correlations of two identical pions at low relative 
momenta were first shown to be sensitive to the spatial scale of 
the emitting source by G. Goldhaber, S. Goldhaber, W. Lee
and A. Pais 50 years ago \cite{GGLP}. The correlation method since developed 
and known at present as ``correlation femtoscopy'' was successfully applied 
to the measurement of the space-time characteristics of particle production 
processes in high energy collisions, especially in heavy-ion collisions 
(see, e.g. \cite{pod89,led04,lis05}).

The main goals for carrying out the study of femtoscopic 
correlations with charged kaons are: 
1) study the transverse mass ($m_{\rm T}$) dependence of the correlation radii,  
2) extend the available $m_{\rm T}$ range,  
3) get strong constraints for hydrodynamic models predictions (the models should work 
for heavier mesons and baryons, not only for pions),
4) get a clearer signal (kaons are less affected 
by the decay of resonances than pions),
5) get a consistency check with ${\rm K}^0_{\rm s}{\rm K}^0_{\rm s}$.  

Previous ${\rm K}^{\rm ch}{\rm K}^{\rm ch}$ studies carried out in 
Pb-Pb collisions at SPS by the NA44 and NA49 Collaborations \cite{KK_SPS} reported the decrease
of the long radii ($R_{long}$) with $m_T$ as $\sim m_{T}^{-0.5}$ as a consequence of 
the boost-invariant longitudinal flow \cite{Sinyukov88}; 
the transverse radii ($R_T$) decrease as  $\sim m_{T}^{-0.3}$.
The common $m_{T}$-scaling for $\pi$ and $K$ is an indication of 
the fact that the thermal freeze-out occurs simultaneously for $\pi$ and $K$ 
and they receive a common Lorentz boost. The weak dependence of $R_T$ 
on $m_T$ was successfully  reproduced by a hydrodynamic model \cite{Chapman95}
with $T$=120~MeV and $\beta_{T}$=0.55.

Next studies carried out in Au--Au collisions at RHIC by the PHENIX
 \cite{Afanasiev:2009ii} and STAR \cite{KKSTAR:2004}
Collaborations shown the flat $m_T$-dependencies for $\pi$ and $K$ radii, 
consequence of freeze-out at the same hyper-surface. Similarly to SPS no 
exact universal $m_T$-scaling for $R_{long}$, $R_{out}$ and $R_{side}$ was observed.
These experimental data were successfully reproduced by the Hydro Kinetic Model (HKM) model \cite{HKM2011}.

The approximate ``scaling'' in transverse mass (the one-dimensional, 
$R_{inv}$ source sizes versus $m_{\rm T}$ for different particle types ($\pi$, K, p...) 
fall on the same curve with an accuracy of $\sim 10 \%$) and is usually considered as 
an additional confirmation of the hydrodynamic type of expansion \cite{lis05}.

It is important to perform the same femtoscopic studies as for the heavy ion collisions with the system created in very high energy pp collisions at LHC energies. 
Event multiplicities reached in 7~TeV pp collisions at the LHC are 
comparable with those measured in peripheral A+A collisions at RHIC, 
making the study of the particle momentum and the particle transverse mass dependencies of the correlation radii an important test of the collectivity in pp collisions.

The previous ${\rm K}^{\rm ch}{\rm K}^{\rm ch}$ studies 
with elementary collisions were performed in a combined data set from  
$\alpha\alpha$, $pp$, and $p\overline{p}$ collisions at ISR by AFS Collaboration 
\cite{KK_ISR}, in e$^+$e$^-$ collisions at LEP by the OPAL and DELPHI Collaborations 
\cite{KK_LEP}, in ep collisions by the ZEUS Collaboration \cite{KK_HERA}. 
Due to statistics limitations, only one-dimensional radii were extracted 
in these experiments, no multiplicity and transverse momentum studies were performed.

The ALICE Collaboration studied the two-kaon correlation radii in pp collisions at
 7~TeV: ${\rm K}^0_{\rm s}{\rm K}^0_{\rm s}$ \cite{Humanic:2011ef} and  ${\rm K}^{\rm ch}{\rm K}^{\rm ch}$
  \cite{KchKch_pp}. The multiplicity and $m_{\rm T}$ dependencies of the radii were studied.

\section{{Kaon femtoscopic analyses of Pb-Pb collisions at $\sqrt{s_{NN}}$=2.76~TeV}}
\label{sec-1}

The dataset of Pb–Pb collisions at $\sqrt{s_{NN}}$=2.76~TeV
used in this analysis consisted of roughly 20 million events.
Tracks were reconstructed with the Time Projection Chamber (TPC) \cite{ALICE1}. 
The kaons were selected in kinematic ranges: $|\eta| < 0.8$
and $0.14<p_T<1.5$\,GeV/$c$. The quality of the track is determined by the $\chi^2/N$
value for the Kalman fit to the reconstructed position of the TPC
clusters ($N$ is the number of clusters attached to the track);
the track was rejected if the value is larger than 4.0.
In order to reduce the number of secondary particles it is
required that the distance from particle trajectory to the primary
vertex (distance of the closest approach, DCA) is less
than 2.4\,cm in the transverse plane and 3.0\,cm 
in the beam direction.

For the charge kaon analysis the good PID is crucially important.
In the present analysis, kaons are selected using
the TPC and TOF detectors.
It is required that the deviation of the specific (d$E$/d$x$) energy loss
in TPC from the one calculated using parametrized Bethe-Bloch
formula is within some number of sigma standard deviations
($N_{\sigma TPC}$). A similar $N_{\sigma TOF}$ method is
applied for the particle identification in TOF. The
difference is considered between the measured time-of-flight and the calculated
one as a function of the track length and the particle momentum for each particle mass.
The following cuts are used for kaon selection in
TPC and TOF: $N_{\sigma TPC}<2$ at $p_T<0.5$~GeV/$c$;
at $p_T>0.5$~GeV/$c$ $N_{\sigma TPC}<3$ and $N_{\sigma TOF}<2$. 

Usually the femtoscopic correlation functions of identical
particles are very sensitive to the two-track reconstruction
effects because the considered particles have close momenta and
close trajectories. Two kinds of two-track effects have been
investigated. The ``splitting'' of the tracks means that one track
is reconstructed as two. The ``merging'' means that two different
tracks are reconstructed as one. The following double track cuts were applied in this analysis:
anti-splitting cut: pairs sharing more than 5$\%$ of
clusters in the TPC are removed; anti-merging cut: pairs should be sufficiently well separated in
the angular distance $\Delta \varphi^{*}>0.017$ and in $\Delta \eta >0.02$
at the radius within TPC equal to 1.6\,m. Here
$\varphi^{*}$ is the modified azimuthal angle $\varphi$ taking into
account bending inside the magnetic field.
The analysis has been performed for two signs of the magnetic
field separately. Pair cuts have been applied in exactly the same way for the same
(signal) and mixed (background) pairs.

The two-particle correlation function $C({\bf p}_1,{\bf p}_2)=A({\bf
p}_1,{\bf p}_2)/B({\bf p}_1,{\bf p}_2)$ is defined as a ratio of
the two-particle distribution in the given event $A({\bf p}_1,{\bf
p}_2)$ to the reference one, constructed by mixing particles from
different events of a given class.

The correlation functions were constructed as a functions of
$q_{inv}= \sqrt{|{\bf q}|^{2} - q_{0}^{2}} $, ${\bf q}={\bf p_1}-{\bf p_2}$,
and $q_0=E_1-E_2$. They were parametrized
assuming Gaussian distribution of particle source
in the pair rest frame (PRF). The fitting was performed using the ``Bowler--Sinyukov'' formula:
\begin{equation}
CF(q_{inv})=1 -\lambda +\lambda K(q_{inv})\left( 1+
\exp{\left(-R_{inv}^{2} q^{2}_{inv}\right)}\right)D(q_{inv}),
\label{eq:CF}
\end{equation}
 Factor $K(q_{inv})$ is the Coulomb function integrated over
a spherical source of a given size in PRF.
The parameters $R_{inv}$ and $\lambda$ describe the size of the
 source and the correlation strength, respectively.
The function $D(q_{\rm inv})$, ``baseline'', takes into account all non-femtoscopic 
correlations, including the long-range correlations due to 
energy-momentum conservation. In analysis of pp collisions the baseline 
was fitted by a standard quadratic polynomial; the PYTHIA (PERUGIA-2011) model was
used to determine the parameters of the polynomial \cite{KchKch_pp}.
In analysis of Pb-Pb collisions: $D(q_{inv})=1$. 

\begin{figure}[!h]
\begin{center}
\includegraphics[width=0.8\textwidth]{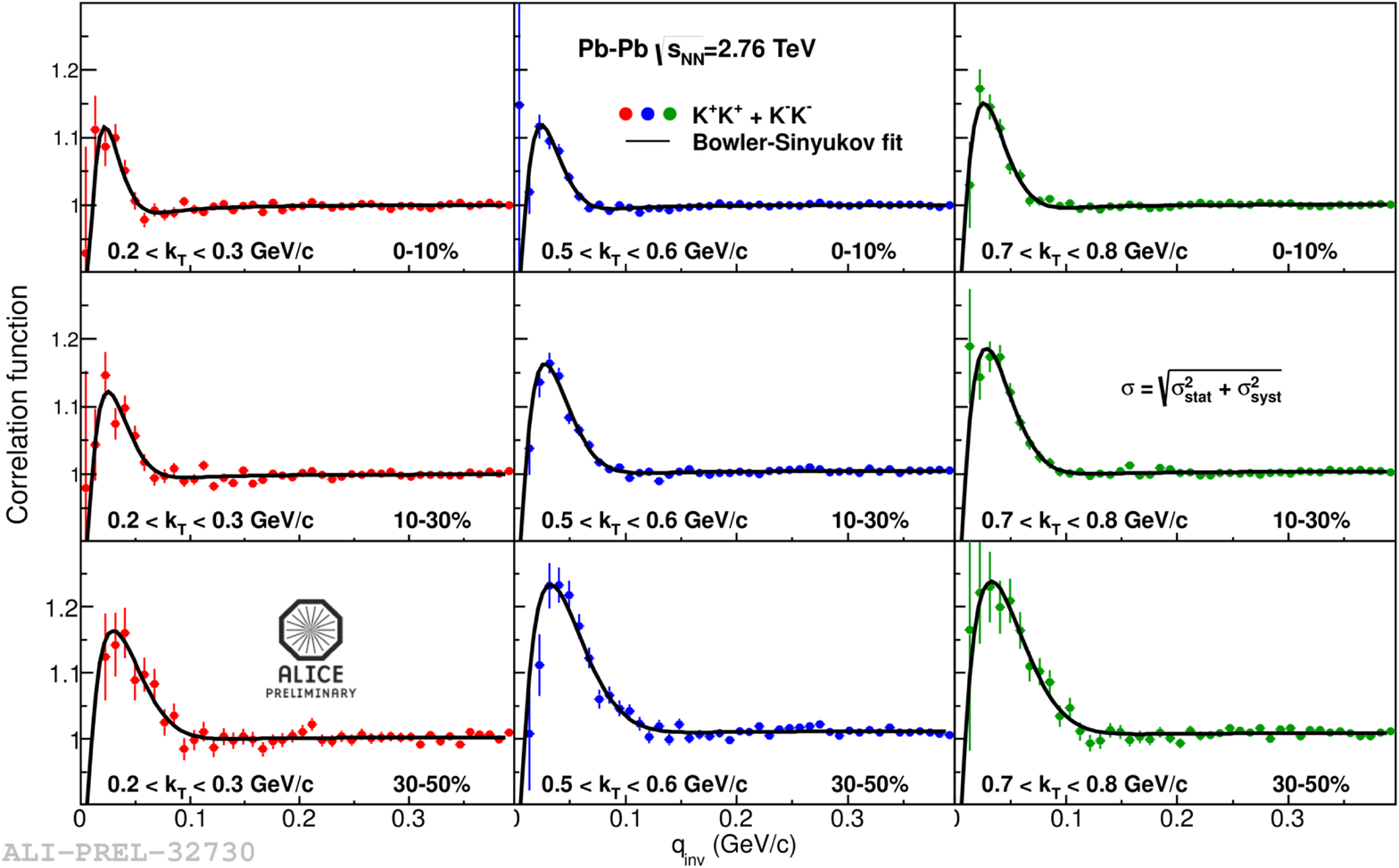}
\caption{
Correlation functions versus $q_{inv}$ for identical kaons from Pb-Pb
collisions at $\sqrt{s_{NN}}=2.76$~TeV (solid circles). Positive and negative
kaon  pairs are combined. The three rows represent 
the samples with different centralities:
 $(0-10)\%$, $(10-30)\%$, $(30-50)\%$, 
the three columns represent the three pair transverse momentum ranges:
(0.2-0.3), (0.5-0.6), (0.7-0.8)~GeV/$c$.
The lines going
through the points represent the Gaussian fits discussed in the text.
}
\label{fig:CFs_KK_PbPb}
\end{center}
\end{figure}

Figure~\ref{fig:CFs_KK_PbPb} presents the experimental two-kaon correlation functions
as a function of the invariant pair relative momentum ($q_{\rm inv}$).
The kaon correlation functions shown in Figure~\ref{fig:CFs_KK_PbPb} 
allows to conclude that radii decrease with increasing $k_T$, the corresponding correlation functions
become wider. The radii are smaller for more
peripheral collisions, the corresponding correlation functions are wider.
The $\lambda$ values are in the range (0.4-0.7), in
agreement with chemical model predictions.

\begin{figure}[!h]
\begin{center}
\includegraphics[width=0.6\textwidth]{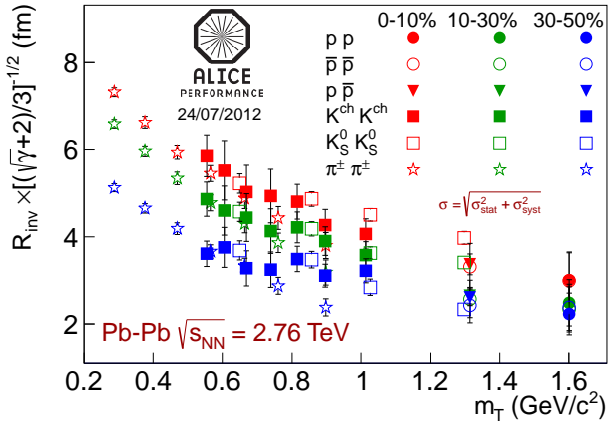}
\caption{ $m_{T}$-dependence of the radius parameter extracted from correlations of pions, 
charged kaons, neutral kaons and protons. Estimates of total errors 
(the quadrature sum of the statistical and the systematic ones) are shown.}
\end{center}
\label{fig:R_piKp}
\end{figure}

The figure~\ref{fig:R_piKp} shows the transverse mass dependence of 
the invariant radii scaled by a kinematic factor for pions, 
kaons and protons \cite{MS_QM2012}. 
As expected, the radii follow the same curve with an accuracy 
$\sim 10\%$ in agreement with hydrodynamic predictions.
.

\section{Kaon femtoscopic analyses of pp collisions at 
$\sqrt{s}$=7~TeV.  }
\label{sec-2}
\begin{figure}[!h]
\begin{center}
\includegraphics[width=0.8\textwidth]{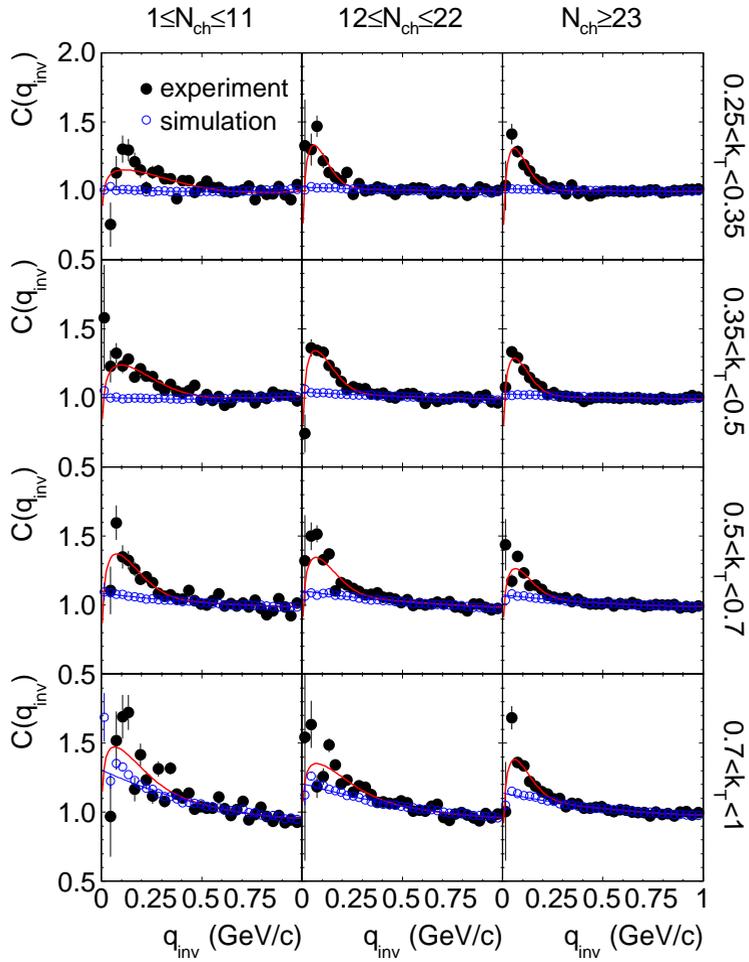}
\caption{
Figure from \cite{KchKch_pp}.
Correlation functions versus $q_{inv}$ for identical kaons from pp
collisions at $\sqrt{s}=7$~TeV (solid circles)
and those obtained with PYTHIA (PERUGIA-2011) (open circles). Positive and negative
kaon  pairs are combined. 
The lines going
through the points represent the Gaussian fits discussed in the text.
}
\label{fig:CFs_KK_pp}
\end{center}
\end{figure}

The dataset of pp collisions at $\sqrt{s}$=7~TeV 
used in this analysis consisted of roughly 300 million minimum-bias events.
The kaons were selected in kinematic ranges: $|\eta| < 1.0$
and $0.15<p_T<1.2$\,GeV/$c$.
The details of analysis can be found in \cite{KchKch_pp}.
Figure~\ref{fig:CFs_KK_pp} presents the experimental two-kaon correlation functions
 and those obtained from a simulation using PYTHIA (PERUGIA-2011) (open circles)
as a function of the invariant pair relative momentum.

\begin{figure}[!h]
\begin{center}
\includegraphics[width=0.49\textwidth]{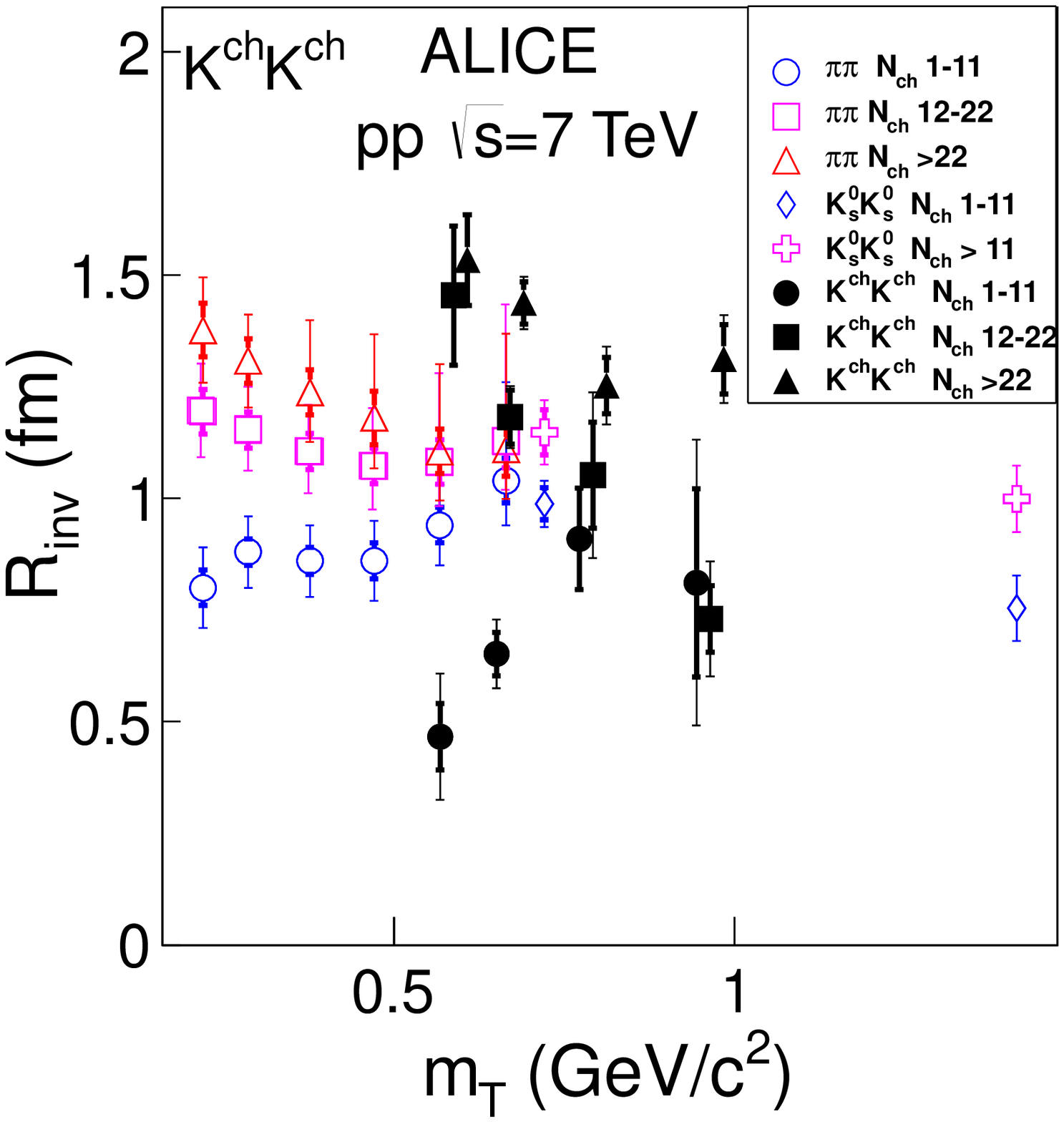}
\includegraphics[width=0.49\textwidth]{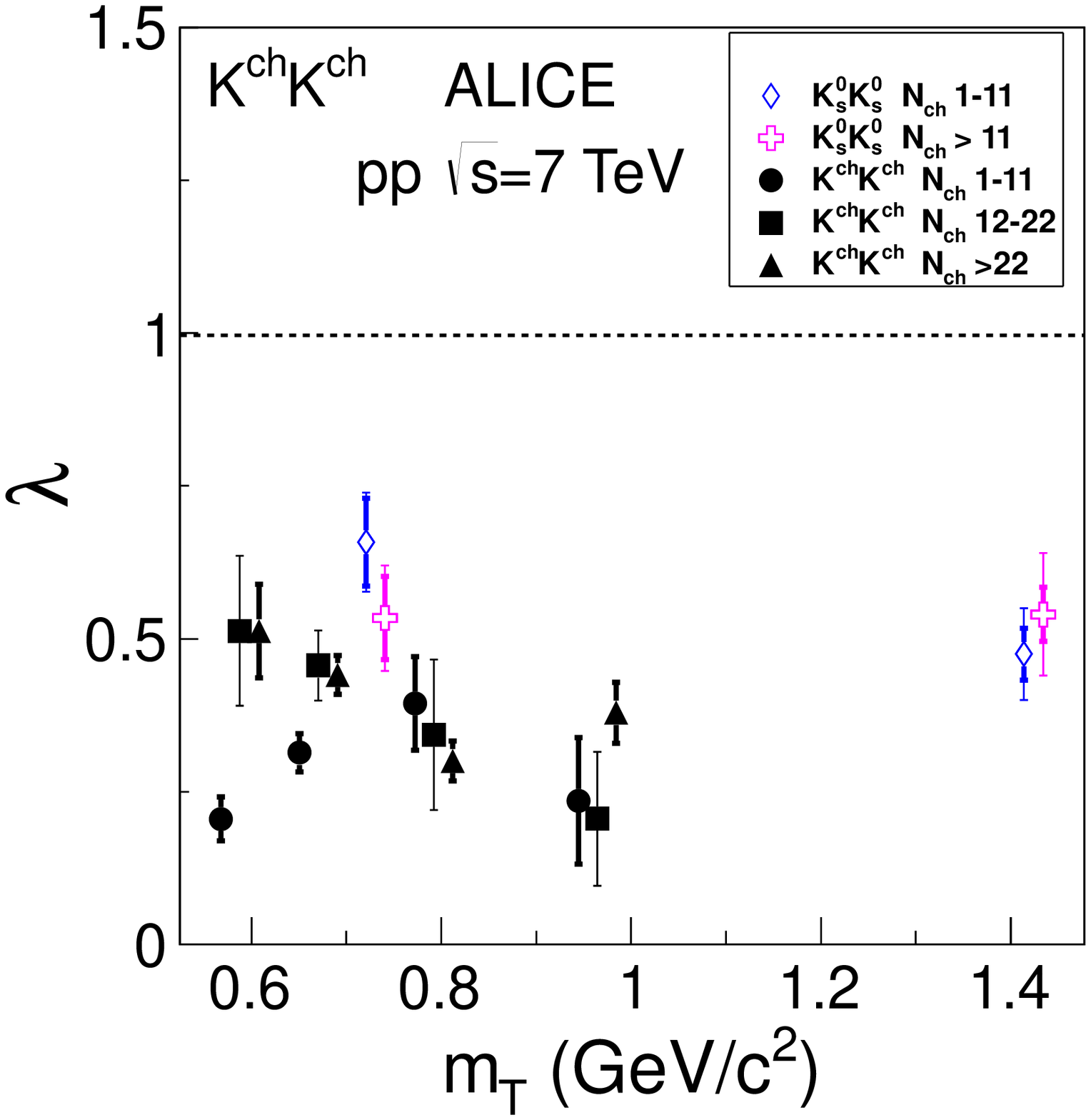}
\caption{ 
Figure from \cite{KchKch_pp}.
Left: 
One-dimensional charged kaon radii versus $m_{\rm T}$.
For comparison the $\pi\pi$\cite{Aamodt:2011kd} 
and  ${\rm K}^0_{\rm s}{\rm K}^0_{\rm s}$ \cite{Humanic:2011ef} 
radii measured by ALICE in 7~TeV
pp collisions are also shown.
Statistical (darker lines) and total errors are shown.
Right:$\lambda$-parameters of ${\rm K}^{\rm ch}{\rm K}^{\rm ch}$
and  ${\rm K}^0_{\rm s}{\rm K}^0_{\rm s}$  vs $m_{\rm T}$  
}  
\label{fig:RLam_pp}
\end{center}
\end{figure}

The ${\rm K}^{\rm ch}{\rm K}^{\rm ch}$ correlation radii in Fig.~\ref{fig:RLam_pp} 
show an increase with multiplicity; these radii also decrease with increasing $m_{\rm T}$ 
for the large multiplicity bins $N_{\rm ch}$ $(12-22)$ and $N_{\rm ch}$ $(>22)$
 as it was observed in heavy-ion collisions.

In the low multiplicity bin $N_{\rm ch}$ $(1-11)$
charged kaons show a completely different $k_{\rm T}$-dependence of the radii:
these radii increase with $k_{\rm T}$. This effect is qualitatively
similar to that of pions \cite{Aamodt:2011kd}.

\section{Summary.  }
\label{sec-3}

In Pb-Pb collisions the charged kaon radii increase with increasing multiplicity and decrease
with increasing pair transverse momentum. The same behavior was observed at large multiplicity in pp collisions.
The femtoscopic radius of pions, kaons and protons in Pb-Pb collisions demonstrate 
an approximate scaling with transverse mass according with hydrodynamic model predictions.
However there is an indication of breaking of such scaling in pp collisions.

\end{document}